\documentclass[final,reqno]{siamltex}
\usepackage{epsfig}
\usepackage{graphicx}
\usepackage{subfigure}
\usepackage{epstopdf}

\def\vb{v_\mathrm{belt}}
\def\vn{v_\mathrm{nozzle}}
\def\se{s_\mathrm{end}}
\def\tm{t}
\def\tme{\tm_\mathrm{end}}
\def\tmew{\tm_\mathrm{end}(w)}
\def\tmewo{\tm_\mathrm{end}(w_1)}
\def\tmewt{\tm_\mathrm{end}(w_2)}
\def\tmewn{\tm_\mathrm{end}(w_n)}

\def\be{\begin{equation}}
\def\ee{\end{equation}}
\def\ts{\tilde{s}}
\def\tse{\tilde{s}_\mathrm{end}}
\def\tv{\tilde{v}}
\def\tvn{\tilde{v}_\mathrm{nozzle}}

\def\vtw{v(\tm;w)}
\def\dvtw{v'(\tm;w)}
\def\vtwo{v(\tm;w_1)}
\def\vtwt{v(\tm;w_2)}
\def\vtwn{v(\tm;w_n)}
\def\tO{\tilde{\Theta}}

\def\B{B}
\def\I{I}
\def\Jtt{J_{\frac{2}{3}}}
\def\Jmtt{J_{-\frac{2}{3}}}
\def\Jot{J_{\frac{1}{3}}}
\def\Jmot{J_{-\frac{1}{3}}}
\def\hzz{\hat{z}_0}
\def\halp{\hat{\alpha}}

\def\es{\mathbf{e}_s(s)}
\def\en{\mathbf{e}_n(s)}

\def\mcA{\mathcal{A}}

\title{Falling of a viscous jet onto a moving surface}

\author{A. Hlod\footnotemark[2]\ \footnotemark[3]\and A.C.T. Aarts\footnotemark[2] \and A.A.F. van de Ven\footnotemark[2]\and  M.A. Peletier\footnotemark[2]}

\begin{document}

\maketitle
\renewcommand{\thefootnote}{\fnsymbol{footnote}}

\footnotetext[2]{Center for Analysis, Scientific computing and Applications, Eindhoven University of Technology, Eindhoven, The Netherlands.}
\footnotetext[3]{Author to whom all correspondence should be addressed. E-mail: a.hlod@tue.nl. Postal address: Dept. of Mathematics and Computer Science, Technische Universiteit Eindhoven, PO Box 513, 5600 MB  Eindhoven, The Netherlands.}

\renewcommand{\thefootnote}{\arabic{footnote}}
\begin{abstract}
We analyze the stationary flow of a jet of Newtonian fluid that is drawn by gravity onto a moving surface. The situation is modeled by a third-order ODE on a domain of unknown length and with an additional integral condition; by solving part of the equation explicitly we can reformulate the problem as a first-order ODE, again with an integral constraint. We show that there are two flow regimes,   and characterize the associated regions in the three-dimensional parameter space in terms of an easily calculable quantity. In a qualitative sense the results from the model are found to correspond with experimental observations.
\end{abstract}



\pagestyle{myheadings} \thispagestyle{plain}
\markboth{ A. HLOD, A.C.T. AARTS, AND A.A.F. VAN DE VEN, M.A. PELETIER}{FALLING JET ONTO MOVING SURFACE}
\section{Introduction} \label{sec:Introd}

\begin{figure}[t]
    \center\scalebox{0.4}{\includegraphics{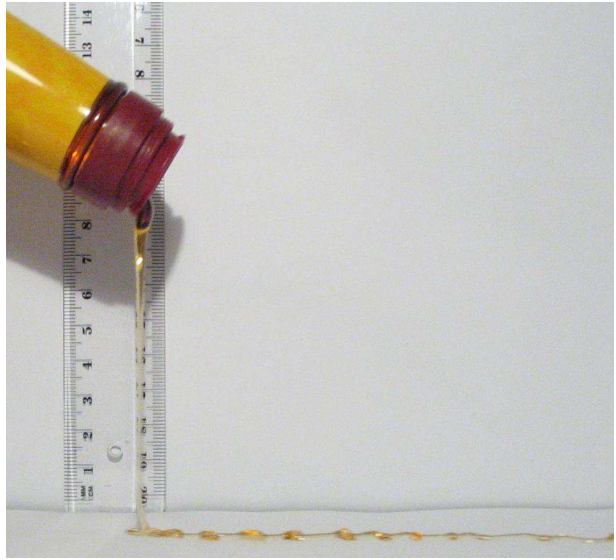}}
   \caption{Straight flow of syrup for low surface velocity and large bottle height.}
   \label{fig:photo_straight}
\end{figure}

\begin{figure}[t]
    \center\scalebox{0.4}{\includegraphics{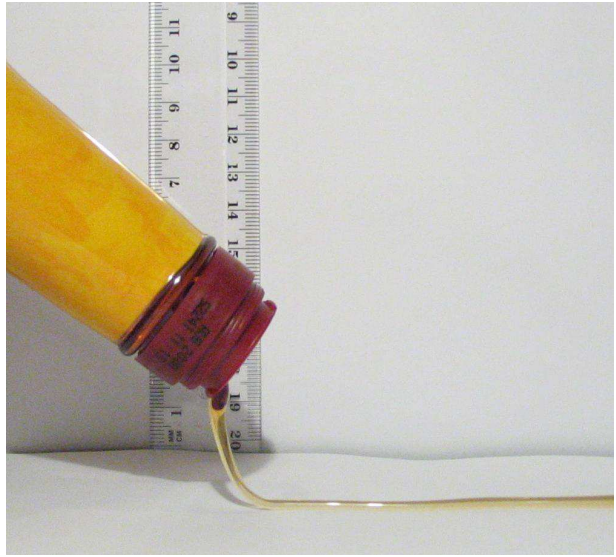}}
   \caption{Curved flow of syrup for high surface velocity and small bottle height.}
   \label{fig:photo_curved}
\end{figure}

In the flow of a viscous fluid jet falling onto a moving surface different flow regimes can be distinguished, as is easily observed if one pours syrup onto a pancake (Figs~\ref{fig:photo_straight} and~\ref{fig:photo_curved}). If the syrup is poured from a large height and the bottle moves slowly, then the syrup thread is perfectly vertical (Fig.~\ref{fig:photo_straight}). If the bottle is held closer to the pancake and moved relatively fast, however, then the flow of syrup becomes curved.
%

Note the difference in behavior at the touchdown point. In the straight case, the jet hits the surface at right angles, and a little puddle forms on the surface from which a thicker `jet' is transported away by the surface movement (Fig.~\ref{fig:photo_straight}). Since the flow velocity of the surface jet equals the surface velocity, conservation of mass implies that at the touchdown point the particle velocity in the free jet is larger than the velocity of the surface. In the curved case, however, the jet does not thicken at touchdown, and the free jet meets the surface tangentially.

There is a large body of literature on viscous jets or sheets that impinge upon \emph{fixed} surfaces, where one can observe folding of viscous sheets \cite{Scorobogatiy,Ribe1}, coiling of viscous jets \cite{Ribe2}, and viscous fluid buckling \cite{Cruickshank}.
The fact that in these cases the surface is stationary is essential; to our knowledge the current work is the first study of a viscous jet that falls upon a \emph{moving} surface.

In Section~\ref{sec:MathMod}, we construct a mathematical model  of the flow, where we first concentrate on the curved jet of Fig.~\ref{fig:photo_curved}. We model the flow as a thin, Newtonian jet of \emph{a priori} unknown length. In Section~\ref{sec:Transf}, the original system of equations is transformed to a first-order differential equation for the flow velocity and two additional relations for two unknown parameters. In Section~\ref{sec:ExistUnique}, we show that in a certain parameter regime the original system admits a unique solution, and we give a convenient characterization of the relevant part of parameter space.
In Section~\ref{sec:numsol}, we present the solution algorithms for the model equations. Results for various model parameters are shown in Section~\ref{sec:results}, and in Section~\ref{sec:concl}, we discuss our results and give some conclusions.
\section{Mathematical model} \label{sec:MathMod}
\begin{figure}[ht]
    \center\scalebox{0.8}{\includegraphics{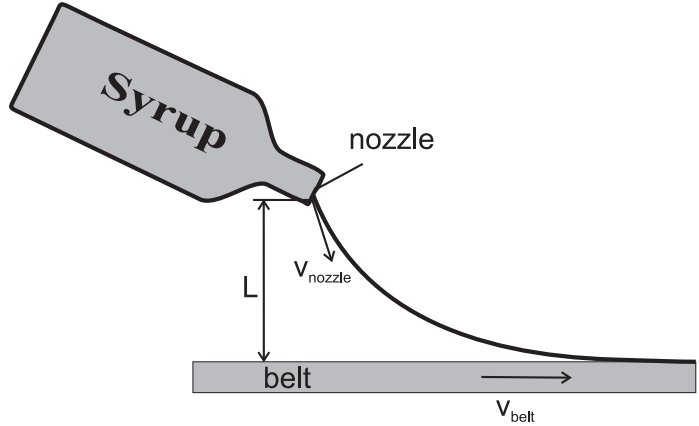}}
   \caption{Jet falling from the nozzle onto the moving belt.}
   \label{Fig3}
 \end{figure}

A thin stream of Newtonian fluid with viscosity~$\eta$ and density~$\rho$ is falling from the nozzle of a bottle onto a moving  belt (Fig.~\ref{Fig3}).
We use the theory  of thin jets (see \emph{e.g.}~\cite{Yarin}) and thus
describe the jet as a curve.
The magnitude of the flow velocity at the nozzle is $\vn$, the belt velocity is $\vb$, and the distance between the nozzle and the belt is $L$. The flow of the fluid is stationary and the jet has a curved shape. We restrict ourselves to curves under tension and therefore require that
\be
\vb>\vn. \label{eq:VbVn}
\ee
In Lemma \ref{lemPropofV}, we show that for a curved jet the flow velocity increases from the nozzle to the belt which justifies (\ref{eq:VbVn}).
\par
\begin{figure}[h]
    \center\scalebox{1}{\includegraphics{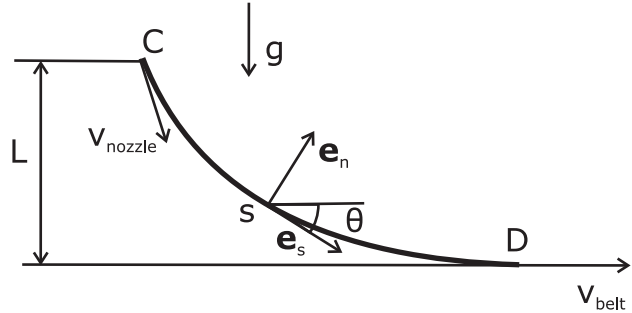}}
   \caption{The geometry of the jet falling from the nozzle onto the moving belt.}
   \label{Fig4}
\end{figure}

The part of the jet between the nozzle and the belt is represented by its center line, the curve $CD$ in Fig.~\ref{Fig4}. The point $C$ indicates the nozzle and the point $D$ indicates the contact with the belt. The acceleration of gravity is $\mathbf{g}$ and the belt surface is perpendicular to $\mathbf{g}$. We parameterize the center line by arclength $s$ ($s=0$ at $C$, and $s=\se$ at $D$). For each point on the curve we define a local orthonormal coordinate system $\es, \en$ consisting of the tangent and normal unit vectors at the point $s$. The angle between $\es$ and the belt surface is $\Theta(s)$. The cross-sectional area at $s$ is $\mcA(s)$, and the average velocity of the fluid at this point is $\mathbf{v}(s)=v(s)\es$.
\par
The flow of  fluid is described by the equations of conservation of mass and balance of momentum~\cite[(4.18)]{Yarin}, which for stationary flow are
\be
 (\mcA(s)v(s))'=0, \label{eq:massbalance}
\ee
\be
(\mcA(s)v(s)\mathbf{v}(s))'=\frac{1}{\rho}(P(s)\es)'+\mathbf{g}\mcA(s), \label{eq:momentumbalance}
\ee
respectively, where by $'$ we denote differentiation with respect to $s$.  The longitudinal force $P(s)$  is obtained from the constitutive law for a Newtonian viscous fluid
\be
    P(s)=\eta_T \mcA(s) v'(s). \label{eq:longforce}
\ee
Here $\eta_T$ is the Trouton elongational viscosity, which for a Newtonian fluid equals $3\eta$ \cite{Yarin}.
\par
Using (\ref{eq:massbalance}) and (\ref{eq:longforce}), we write the balance of momentum (\ref{eq:momentumbalance}) in components in the coordinate system $\es, \en$, as
\begin{eqnarray}
v'(s) =\frac{g \sin \Theta(s)}{v(s)}+
\mu\left( \frac{v'(s)}{v(s)}\right)', \label{ME1_1}
\\
v(s)\Theta'(s)  =\frac{g \cos \Theta(s)}{v(s)}+
\mu\left( \frac{v'(s)}{v(s)} \right)\Theta'(s), \label{ME1_2}
\end{eqnarray}
where $\mu$ is equal to three times the kinematic viscosity, i.e. $\mu=3\eta/\rho$.
\par
Since system (\ref{ME1_1}--\ref{ME1_2}) is of second order with respect to the velocity $v(s)$ and of first order with respect to the angle $\Theta(s)$, we need two boundary conditions for $v(s)$ and one for $\Theta(s)$.  For $v(s)$ we know the velocity of the jet at the nozzle (point $C$) and the velocity at the contact with the belt (point $D$)
\begin{eqnarray}
v(0)=\vn, \label{BC1}
\\
v(\se)=\vb. \label{BC2}
\end{eqnarray}
Note that the length $\se$ of the jet $CD$ is unknown. The angle $\Theta(s)$ at the contact with the belt is zero, so
\be
\Theta(\se)=0. \label{BC3}
\ee
Because the length of the belt $\se$ is unknown in advance, we need an additional condition relating $\se$ to the distance $L$ between the nozzle and the belt
\be
L=\int_0^{\se}\sin \Theta(s)\, ds. \label{CC4}
\ee
The equations (\ref{ME1_1}) and (\ref{ME1_2}) together with the three boundary conditions (\ref{BC1}--\ref{BC3}), and the additional condition (\ref{CC4}) form the complete system for the unknowns~$v(s)$, $\Theta(s)$ and $\se$.
\par
We next make the equations dimensionless. We scale the length $\se$ with respect to $\mu/\vb$, reverse the direction of $s$, and move the origin of $s$ to the point $D$, i.e. $\ts := \vb(\se-s)/\mu$. The velocity $v(s)$ is scaled with respect to the velocity of the belt $\vb$, i.e., $\vb\tv(\ts):=v(s)$. Also we introduce a new angle $\tO(\ts):=\Theta(s)$. The scaled version of (\ref{ME1_1}--\ref{CC4}) reads
\begin{eqnarray}
\left( \tv(\ts)+\frac{\tv'(\ts)}{\tv(\ts)}\right)'=- A \frac{
\sin\tO(\ts)}{\tv(\ts)}, \label{DLE1}
\\
\tO'(\ts)  =-A\frac{ \cos \tO(\ts)}{\tv(\ts)\left(
\tv(\ts)+\frac{\tv(\ts)'}{\tv(\ts)}\right)}~,\label{DLE2}
\\
\tv(0)=1, \label{DLE3}
\\
\tv(\tse)=\tvn,\label{DLE4}
\\
\tO(0)=0, \label{DLE5}
\\
\int_0^{\tse}\sin \tO(\ts)\,d\ts=\B. \label{DLE6}
\end{eqnarray}
Here
\[
A=\frac{g\mu}{\vb^3},\qquad  \B=\frac{\vb L}{\mu},
  \qquad  \tvn=\frac{\vn}{\vb}, \qquad\mbox{and}\qquad
  \tse=\frac{\se\vb}{\mu}.
\]
All these parameters are positive and $\B$ is the Reynolds number. The prime now denotes differentiation with respect to $\ts$, and in the sequel we omit the tildes.
\section{A first-order differential equation for the velocity} \label{sec:Transf}
By introducing a new variable $\xi(s)$
\be
 \xi(s)=v(s)+\frac{v'(s)}{v(s)}, \label{ExrForXi}
\ee
we can rewrite the equations (\ref{DLE1}--\ref{DLE2}) as
\begin{eqnarray}
v(s)\xi'(s)=-A \sin \Theta(s), \label{xiseq1}
\\
v(s)\Theta'(s)  =-A\frac{ \cos \Theta(s)}
 {\xi(s)}, \label{xiseq2}
\\
 v'(s)=\xi(s)v(s)-v^2(s). \label{xiseq3}
\end{eqnarray}
For the variable $\xi(s)$ it is necessary to provide an initial value. To compute it, we need to know values of $v(s)$ and $v'(s)$ at the same point. Because we do not know a value of $v'(s)$ at any point we prescribe a value of $\xi(s)$ at $s=0$,
\be
 \xi(0)=-\sqrt{w}, \ w\ge 0. \label{IncondXi}
\ee
Here we restrict ourselves to a negative initial value for $\xi$. Further in this section, see (\ref{SolXiThetaStep4}) and (\ref{SolXiThetaStep5}), we explain our choice of the form for the initial value for $\xi(s)$. The value $w$ is unknown in advance and is determined by the requirement that a solution of (\ref{xiseq1}--\ref{xiseq3}) has to satisfy the conditions (\ref{DLE6}) and (\ref{DLE4}).
\par
Next, we replace the material coordinate $s$ by the time $\tm$, according to
 $$
 ds= v(\tm) d\tm,
 $$
 in the system of equations (\ref{xiseq1}--\ref{xiseq3}) together with the conditions (\ref{DLE3}--\ref{DLE6}) and (\ref{IncondXi}), and we obtain
\begin{eqnarray}
\xi'(\tm)=-A \sin \Theta(\tm), \label{Taueq1}
\\
\Theta'(\tm)  =-A\frac{ \cos \Theta(\tm)}
 {\xi(\tm)}, \label{Taueq2}
\\
 v'(\tm)=\xi(\tm)v^2(\tm)-v^3(\tm), \label{Taueq3}
 \\
\xi(0)=-\sqrt{w},\label{Taueq4}
\\
\Theta(0)=0, \label{Taueq5}
\\
v(0)=1,\label{Taueq6}
\\
v(\tme)=\vn,\label{Taueq7}
\\
\int_0^{\tme}v(\tm)\sin \Theta(\tm)\,d\tm=\B. \label{Taueq8}
\end{eqnarray}
Here
$$\tme= \int_0^{\se} \frac{ds}{v(s)}$$
represents the dimensionless time necessary to flow from the nozzle to the belt, which is unknown in advance.
\par
To solve the equations (\ref{Taueq1}) and (\ref{Taueq2}) we multiply (\ref{Taueq1}) by $\sin\Theta(\tm)$ and (\ref{Taueq2}) by $\cos \Theta(\tm)\xi(\tm)$ and add them, to obtain
\be
  (\sin \Theta(\tm)\xi(\tm))'=-A. \label{SolXiThetaStep1}
\ee
We  integrate (\ref{SolXiThetaStep1}) with respect to $\tm$ and use the initial condition (\ref{Taueq5}) to obtain
\be
  \xi(\tm)\sin \Theta(\tm)=-A\tm. \label{SolXiThetaStep2}
\ee
By eliminating $\sin \Theta(\tm)$ from (\ref{SolXiThetaStep2}) and substituting it into (\ref{Taueq1}) we derive the differential equation for $\xi(\tm)$
\be
\xi'(\tm)=\frac{A^2\tm}{\xi(\tm)}, \label{SolXiThetaStep3}
\ee
which has the solution
\be
\xi(\tm)=\pm\sqrt{A^2\tm^2+\xi(0)^2}. \label{SolXiThetaStep4}
\ee
Here we have to choose a correct branch of the square root (\ref{SolXiThetaStep4}). The branch with the positive sign gives negative  $\sin \Theta(\tm)$, see (\ref{SolXiThetaStep2}), which implies an upward-sloping jet; the physically reasonable choice is therefore the branch with the negative sign. With the initial condition (\ref{IncondXi}) we get
\be
\xi(\tm)=-\sqrt{A^2\tm^2+w}, \label{SolXiThetaStep5}
\ee
and from (\ref{SolXiThetaStep2}) and (\ref{SolXiThetaStep5}) we find
\be
\Theta(\tm)=\arcsin \frac{A\tm}{\sqrt{A^2\tm^2+w}}. \label{SolXiThetaStep6}
\ee
\par
Summarizing, the problem (\ref{Taueq1}--\ref{Taueq8}) simplifies to
\begin{eqnarray}
v'(\tm)=-v^2(\tm)(\sqrt{A^2\tm^2+w}+v(\tm)),\label{EqForV(t)1}
\\
v(0)=1,\label{EqForV(t)2}
\\
v(\tme)=\vn,\label{EqForV(t)3}
\\
\int_0^{\tme}\frac{A\tm v(\tm)}{\sqrt{A^2\tm^2+w}}\,d\tm=\B. \label{EqForV(t)4}
\end{eqnarray}
 The unknowns of the problem (\ref{EqForV(t)1}--\ref{EqForV(t)4}) are the velocity $v(\tm)$ and the two positive parameters $w$ and $\tme$.
\section{Existence and uniqueness} \label{sec:ExistUnique}
We reformulate the problem (\ref{EqForV(t)1}--\ref{EqForV(t)4}) as an algebraic equation for the parameter $w$. First we formulate properties of a solution $v(\cdot,w)$ of (\ref{EqForV(t)1}--\ref{EqForV(t)2}) for given $w\ge0$ .
\begin{lemma}
\label{lemPropofV} For any $w\ge0$, equation (\ref{EqForV(t)1}) has a unique solution $v(\cdot,w):[0,\infty)\rightarrow(0,1]$ satisfying (\ref{EqForV(t)2}) with $v(\cdot,w)\in C^1([0,\infty))$.
\par
In addition,
\begin{remunerate}
\item \label{monofv} $\vtw$ is a strictly decreasing function of $\tm$ for fixed $w$ and a strictly decreasing function of $w$ for fixed $\tm$.
\item \label{boundofv}
\be
    \vtw< \frac{2 }{2+\tm \sqrt{A^2 \tm ^2+w}}. \label{eq:UpEstOfV}
\ee
\item
\label{contofvw} The operator $w\mapsto v(\cdot,w)$ is continuous from $[0,\infty)$ to $L^\infty(0,\infty)$.
\end{remunerate}
\end{lemma}
\begin{proof}
The right-hand side of (\ref{EqForV(t)1}) is $C(\Omega)$ and Lipshitz continuous in $v$ uniformly on $\Omega$, where $\Omega=(\{\tm,v,w\}: \tm\in [0,\infty), v\in(0,1], w\ge0)$. Therefore, locally there exists a unique solution of (\ref{EqForV(t)1}) satisfying (\ref{EqForV(t)2}), which continuously depends on $w$ \cite[Theorem 7.4]{Coddington}.
\par
From (\ref{EqForV(t)1}) it follows that $\dvtw<0$ whenever $\vtw>0$ and that $v\equiv0$ is a solution of this equation. Thus, because of (\ref{EqForV(t)2}), $\dvtw$ is always negative and $v(\cdot;w)$ is strictly decreasing. Since $v\equiv0$ is a solution of (\ref{EqForV(t)1}), $\vtw$ remains positive for $\tm\ge0$. Therefore, $\vtw \in (0,1]$ $\forall \tm\ge0$; this proves the existence and uniqueness of $v$ and the monotonicity in $t$.
\par
For the monotonicity in $w$, fix $w_1>w_2\ge0$. Then $v'(0;w_1)=-(\sqrt{w_1}+1)<v'(0,w_2)=-(\sqrt{w_2}+1)$, and $\vtwo<\vtwt$ for small $\tm>0$. Suppose that there exists a $\tm^*>0$ such that $v(\tm^*;w_1)=v(\tm^*;w_2)$; then $v'(\tm^*;w_1)\ge v'(\tm^*;w_2)$, which  leads to a contradiction  with $w_1\le w_2$. This completes the proof of part \ref{monofv} of the Lemma.
\par
Because $\vtw>0$ we have
$$
 \dvtw< -\vtw^2\sqrt{A^2\tm^2+w},
$$
or
\be
 \left(\frac{1}{\vtw}\right)' > \sqrt{A^2\tm^2+w}. \label{eq:ineqL1P2}
\ee
We integrate (\ref{eq:ineqL1P2}) from $0$ to $\tm$ and apply the initial condition $v(0;w)=1$ to find the following estimate of $\vtw$:
$$
\vtw < \frac{2 A}{2A+A\tm \sqrt{A^2 \tm ^2+w}+w \log
\left(\frac{A \tm
   +\sqrt{A^2 \tm ^2+w}}{\sqrt{w}}\right)} <
   \frac{2 }{2+\tm \sqrt{A^2 \tm ^2+w}}.
$$
This estimate proves part~\ref{boundofv} and shows that $\vtw\rightarrow 0$ as $\tm\rightarrow \infty$.
\par
The right-hand side of (\ref{EqForV(t)1}) depends continuously on $w$. This together with the estimate (\ref{eq:UpEstOfV}) of $\vtw$ at $t=\infty$ proves~\ref{contofvw}.
\end{proof}
\par
In order to solve (\ref{EqForV(t)1})-(\ref{EqForV(t)4}) we need to find $w$ for which (\ref{EqForV(t)3})-(\ref{EqForV(t)4}) are satisfied. Knowing a correct value of $w$, we can obtain a solution $v(\tm)$ which leads to a solution of the original problem (\ref{DLE1})-(\ref{DLE6}). Therefore, next we concentrate on finding a correct $w$.
\begin{definition} \label{def:I}
We define a function $\I:[0,\infty)\rightarrow[0,\infty)$ in the following way. For given $w\in [0,\infty)$ let $v(\cdot,w)$ be the solution of (\ref{EqForV(t)1}--\ref{EqForV(t)2}) given by Lemma \ref{lemPropofV}. By items \ref{monofv} and \ref{boundofv} of Lemma \ref{lemPropofV} there exists a unique $\tmew\ge0$ satisfying
\be
v(\tmew;w)=\vn. \label{eq:defoftew}
\ee
Define $\I(w)$ as
\be
     \I(w)=\int_0^{\tmew}\frac{A\tm \vtw}{\sqrt{A^2\tm^2+w}}\,d\tm. \label{defofI}
\ee
\end{definition}%
By Lemma \ref{lemPropofV}, part \ref{boundofv} the integrable function is bounded from above and the integral converges.
\begin{corollary}
\label{cor:existenceI=w}
Solving (\ref{EqForV(t)1}--\ref{EqForV(t)4}) is equivalent to finding a $w\geq0$ that satisfies
\be
    \I(w)=\B. \label{eq:for_w}
\ee
\end{corollary}%
In the next three lemmas we will show some properties of $\I(w)$ which lead to  a characterization of existence and uniqueness of a solution to (\ref{eq:for_w}).
\begin{lemma}
\label{lemMonotI(w)} $\I(w)$ is a strictly decreasing function of $w$.
\end{lemma}
\begin{proof}
Choose $w_1$ and $w_2$ with
\be
 w_1>w_2\ge0. \label{w1w2}
\ee
From part \ref{monofv} of Lemma \ref{lemPropofV} it follows that
\be
 \tmewo<\tmewt. \label{IneqT_end}
\ee
Combining  (\ref{IneqT_end}) with the statement \ref{monofv} of Lemma \ref{lemPropofV} and (\ref{w1w2}) with the definition of $\I(w)$, we have
\begin{eqnarray}
\I(w_1)= \int_0^{\tmewo}\frac{A \tm
 \vtwo}{\sqrt{A^2\tm^2+w_1}}\,d\tm
&<&\int_0^{\tmewt}\frac{A \tm \vtwo}{\sqrt{A^2\tm^2+w_1}}\,d\tm \nonumber \\
&<& \int_0^{\tmewt}\frac{A \tm \vtwt}{\sqrt{A^2\tm^2+w_2}}\,d\tm  =\I(w_2), \nonumber
\end{eqnarray}
   which proves the Lemma.
\end{proof}
\begin{lemma}
\label{lemContI(w)} $\I(w)$ is continuous.
\end{lemma}
\begin{proof}
Fix $w\ge0$ and let
\be
    w_n\rightarrow w\quad \mbox{as}\quad  n \rightarrow \infty. \label{eq:wn}
\ee
Then
 \begin{eqnarray}
    \I(w)-\I(w_n)&=& \int_0^{\tmew}\frac{A \tm
                     \vtw}{\sqrt{A^2\tm^2+w}}\,d\tm
               -
                   \int_0^{\tmewn}\frac{A \tm
                     \vtwn}{\sqrt{C^2\tm^2+w_n}}\,d\tm
                     \nonumber \\
              &=& \int_0^{\tmewn}\left[\frac{A \tm
                     \vtw}{\sqrt{A^2\tm^2+w}}-\frac{A \tm
                     \vtwn}{\sqrt{A^2\tm^2+w_n}}\right]\,d\tm
               +
                   \int_{\tmewn}^{\tmew}\frac{A \tm
                     \vtw}{\sqrt{A^2\tm^2+w}}\,d\tm
                     \nonumber \\
                &=& J_1+J_2. \nonumber
\end{eqnarray}
Both $J_1$ and $J_2$ converge to zero as $n \rightarrow \infty$; for $J_1$ this follows from the continuity of $\vtw$ in $w$ (Lemma \ref{lemPropofV}) and for $J_2$ from the continuity of $\tmew$ in $w$, which we prove next.
\par
From Lemma \ref{lemPropofV} we have that $v(\cdot;w)\in C^1([0,\infty))$ and $-\infty<v_\tm(\tm;w)<0$. Therefore, by the Inverse Function Theorem (\emph{e.g.} \cite[Theorem 9.24]{MathAn}) there exists a function $\tm=\tm(\cdot;w)\in C^1((0,1])$ such that $t\bigl(v(\tilde t;w)\bigr) = \tilde t$ for all $\tilde t\geq0$.
\par
Next note that
\be
 v_n:=v(\tmewn;w)\longrightarrow \vn \quad\mathrm{as}  \quad n \rightarrow \infty,\label{eq:vn}
\ee
since
\begin{eqnarray*}
|v(\tmewn;w) - \vn| &=& |v(\tmewn;w) - v(\tme(w_n);w_n)|\\
&\leq& \|v(\cdot;w)-v(\cdot;w_n)\|_\infty \longrightarrow 0
\end{eqnarray*}
by part~\ref{contofvw} of Lemma \ref{lemPropofV}. Therefore, by continuity of  $\tm(\cdot;w)$ we have
$$
\tmewn =\tm\bigl(v(\tmewn;w);w\bigr)=\tm(v_n;w)\longrightarrow  \tm(\vn;w)=\tmew,
$$
which completes the proof.
\end{proof}
\begin{lemma} \label{lemLimI(w)}
$
\lim_{w \rightarrow \infty}I(w)=0.
$
\end{lemma}
\begin{proof}
From the definition of $\I(w)$ and  $\vtw\in(0,1]$ (Lemma \ref{lemPropofV}) we have
\begin{eqnarray}
\I(w)&=& \int_0^{\tmew}\frac{A \tm
\vtw}{\sqrt{A^2\tm^2+w}}\,d\tm
    < \int_0^{\tmew}\frac{A \tm
    }{\sqrt{A^2\tm^2+w}}\,d\tm  \nonumber\\[2\jot]
    &=&  \frac{\sqrt{w+A^2\tmew^2}-\sqrt{w}}{A}
    =  \frac{A\tmew^2}{\sqrt{w+A^2\tmew^2}+\sqrt{w}}. \nonumber
\end{eqnarray}
Because $\tmew$ decreases in $w$, by letting $ w \rightarrow \infty$ we find
$$
 \lim_{w \rightarrow \infty}\I(w)= 0.
$$
\end{proof}

%
%
%
%
Summarizing the results of previous lemmas, we formulate a theorem of existence and uniqueness of a solution to the original problem (\ref{DLE1}--\ref{DLE6}).
\begin{theorem}
\label{th:ExistenceAndUniquness} There exists a solution to the problem
(\ref{DLE1}--\ref{DLE6}) if and only if
\be
    \I(0;A,\vn)>\B. \label{eq:exisCond}
\ee
If it exists, the solution is unique.
\end{theorem}

The theorem follows simply from Lemmas~\ref{lemMonotI(w)}, \ref{lemContI(w)}, and \ref{lemLimI(w)}.
\begin{figure}[ht]
   \center\scalebox{0.8}{\includegraphics{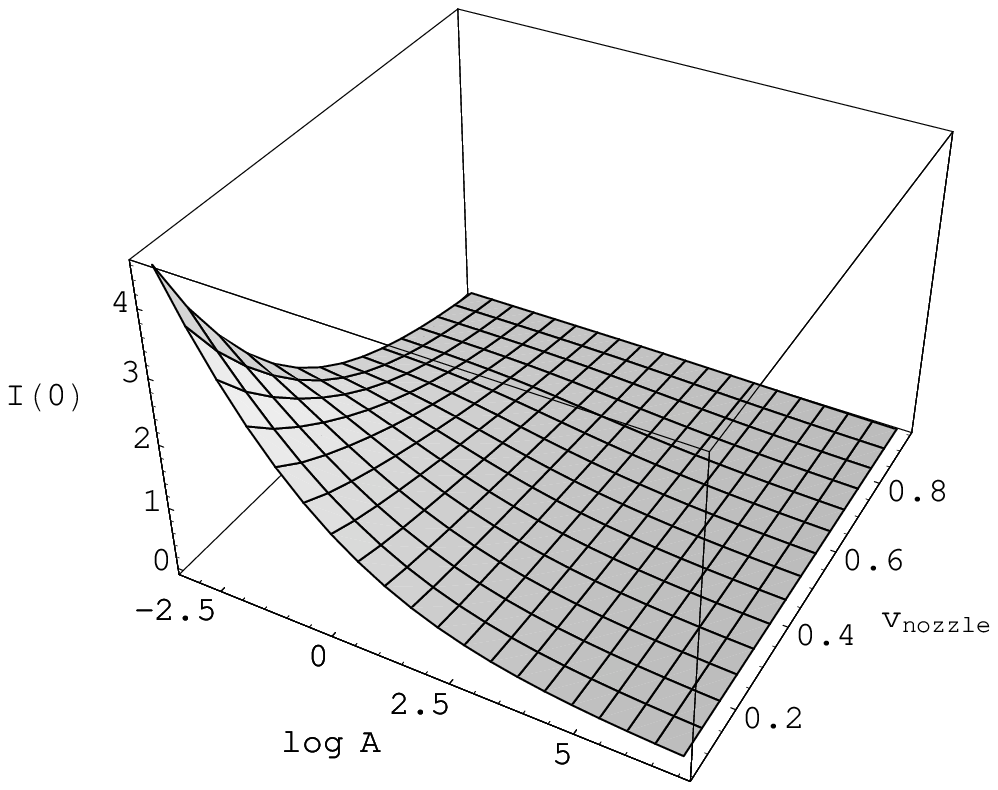}}
   \caption{ Surface $\I(0;A,\vn)$.}
    \label{Fig6}
\end{figure}

\medskip
 In Section \ref{sec:numsol}, we describe two algorithms for computing $\I(0;A,\vn)$ as a function of~$A$ and~$\vn$, resulting in the graph of Fig.~\ref{Fig6}. As a consequence of Theorem \ref{th:ExistenceAndUniquness}, a solution to the original problem (\ref{DLE1}--\ref{DLE6}) exists only if the point $(A,\vn,\B)$ is below the surface $\I(0;A,\vn)$.
\medskip
\par
\textbf{Note: Non-existence of a curved-jet solution.} When the condition (\ref{eq:exisCond}) is not satisfied, then there exists no solution to the curved-jet equations~(\ref{DLE1}--\ref{DLE6}). What happens to a viscous jet in this parameter range is not clear. It is possible that a stable straight jet exists ($\Theta\equiv\pi/2$); close to the transition between existence and non-existence the curved-jet solution becomes straight (see Fig.~\ref{fig7}), supporting this possibility. Other possibilities are coiling (in three dimensions) or buckling (in two dimensions), possibly coexisting with a steady straight jet that is dynamically unstable.

At this moment the issue is open, in part since it is unclear which boundary condition at the lower end of the jet best reflects the physical situation. We plan to return to this question in a future publication.
\section{Numerical approximation} \label{sec:numsol}
The problem of this paper gives rise to two slightly different numerical questions. The first question arises in making a phase diagram such as
Fig.~\ref{Fig6:1}: in order to distinguish between existence and non-existence of a curved jet we need to calculate $\I(0;A,\vn)$ and check the existence condition $\I(0;A,\vn)>\B$ (\ref{eq:exisCond}). The second question arises when this condition is fulfilled: by Corollary~\ref{cor:existenceI=w} we then need to find $w>0$ such that $\I(w;A,\vn)=\B$, from which $v$ and $\Theta$ can then be determined by solving~(\ref{EqForV(t)1}--\ref{EqForV(t)2}) and using (\ref{SolXiThetaStep6}).

The main differential equation~(\ref{EqForV(t)1}) can be solved either analytically or numerically, giving rise to two different methods.
\par
{\bf Method 1.}
When $w=0$, it is possible to solve the problem (\ref{EqForV(t)1}--\ref{EqForV(t)2}) analytically (Appendix~\ref{sec:Appendix}). The rescaled domain size $z^*$ is then to be determined implicitly from
\be
 \vn=\frac{(2A)^{1/3}}{(3z^*)^{2/3}}\left(1+\left(\frac{\Jtt(z^*)c_1
-\Jmtt(z^*)} {\Jot(z^*) +\Jmot(z^*)c_1}\right)^2\right)^{-1}, \ \
c_1=\frac{\Jmtt(\sqrt{2A}/3)}{\Jtt(\sqrt{2A}/3)},
\label{SolforZstar_SN}
\ee
where the $J_\alpha$ are the Bessel functions of the first kind. We then calculate $\tme(0)$ and $\I(0;A,\vn)$ as
\be
\tme(0)=\frac{(6z^*)^{1/3}}{A^{2/3}}
 \frac{\Jtt(z^*)c_1
-\Jmtt(z^*)} {\Jot(z^*) +\Jmot(z^*)c_1},
\ee
\be
 \I(0;A,\vn)=\frac{1-\vn}{\vn}-A\frac{\tme(0)^2}{2}.
\ee
This method only is available for the special case $w=0$.
\par

{\bf Method 2.} Alternatively, one may integrate~(\ref{EqForV(t)1}) numerically until the condition $v(\tm;0)=\vn$ is reached. The integral $\I(w;A,\vn)$ can be computed numerically as well. This method is available for all $w\geq0$.

We solve~(\ref{eq:for_w}) by the bisection method, supplemented with an upper bound on~$w$ that follows from the estimate~(\ref {eq:UpEstOfV}): since for all $w$,
\[
v(t;w) < \frac{2}{At^2},
\]
we have
\[
\vn = v(\tmew;w) < \frac2{A\tmew^2},
\]
and therefore $A\tmew^2 < 2/\vn$. We thus estimate
\be
\I(w;A,\vn)=
\int_0^{\tmew}\frac{A\tm \vtw}{\sqrt{A^2\tm^2+w}}\,d\tm<
\frac{A \tmew^2}{2 \sqrt{w}}<
\frac1{\vn\sqrt w}. \label{eq:estimateofI2}
\ee
Therefore the solution $w$ of~(\ref{eq:for_w}) satisfies  the \emph{a priori} estimate
\[
w\leq \frac1{\vn^2B^2}.
\]
\section{Results} \label{sec:results}
From Theorem \ref{th:ExistenceAndUniquness} it follows that if the parameters $A$, $\B$, and $\vn$ satisfy (\ref{eq:exisCond}), then there exists a solution to the stationary curved-jet equations~(\ref{ME1_1}--\ref{CC4}) (or equivalently~(\ref{EqForV(t)1}--\ref{EqForV(t)4})); otherwise the jet is vertical, or a stationary solution does not exist.
\begin{figure}[ht]
   \center\scalebox{0.7}{\includegraphics{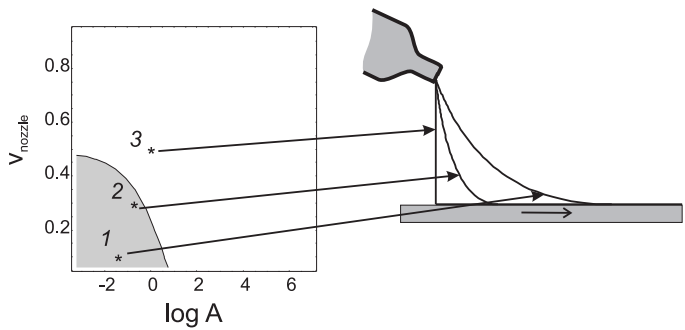}}
   \caption{Region of existence of a solution to the original problem for $\B=1$, (grey region). If a point $(\log A,\,\vn)$ is inside the grey region a curved-jet solution exists (points $1$ and $2$); in point $3$ no such solution exists, and we conjecture that an actual jet is straight. If a point $(\log A,\,\vn)$ is closer to the border of the grey region a shape of the jet is more vertical (point $2$).}
    \label{Fig6:1}
\end{figure}
Fig.~\ref{Fig6:1} shows the region of existence of such a curved-jet solution.
\par
As a reference configuration for the numerical experiments shown below we consider syrup with viscosity $\eta=3.2\,\mathrm{Pa\, s}$  and density $\rho=1000\,\mathrm{kg/m^3}$ pouring from the height $L=2\cdot10^{-2}\,\mathrm{m}$. The velocities of the belt and the flow at the nozzle are $\vb=0.5\,\mathrm{m/s}$ and $\vn=0.05\,\mathrm{m/s}$, respectively.
\begin{figure}[ht]
   \center\scalebox{0.8}{\includegraphics{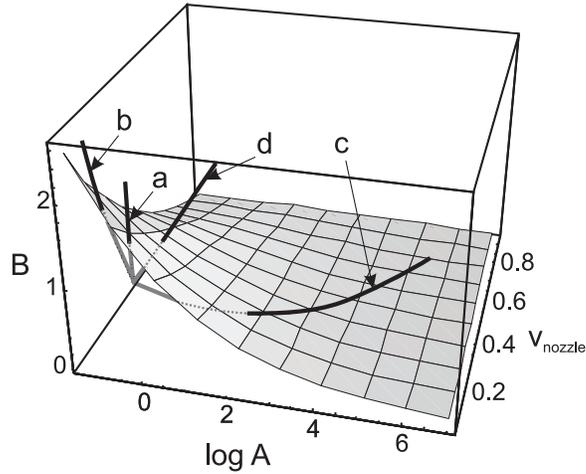}}
   \caption{Curves in non-dimensional parameter space $(A,\vn,\B)$ as we change one of the process parameters $(L,\mu,\vb,\vn)$. The grey parts of the curves below the surface $\I(0;A,\vn)$ correspond to the curved jet; we conjecture that the black parts of the traces correspond to a vertical jet. Line $a$:\  increasing $L$; line $b$:\ decreasing $\mu$; line $c$:\ decreasing $\vb$; and line $d$:\ increasing $\vn$.}
    \label{Fig8}
\end{figure}

Fig.~\ref{Fig8} shows curves in non-dimensional parameter space $(A,\vn,\B)$ corresponding to variation of a single (dimensional) physical parameter  $L$, $\mu$, $\vb$, or $\vn$. In the figure, we see that if $\mu$ or $\vb$ decreases, or if $L$ or $\vn$ increases, the point $(A,\vn,\B)$ eventually leaves the region $\{(A,\vn,\B):\I(0;A,\vn)>\B\}$. Close to this transition the curved jet becomes vertical.
\begin{figure}[h]
\centering
  \subfigure[Varying $L$.]
  {\label{fig7:1}\scalebox{0.6}{\includegraphics{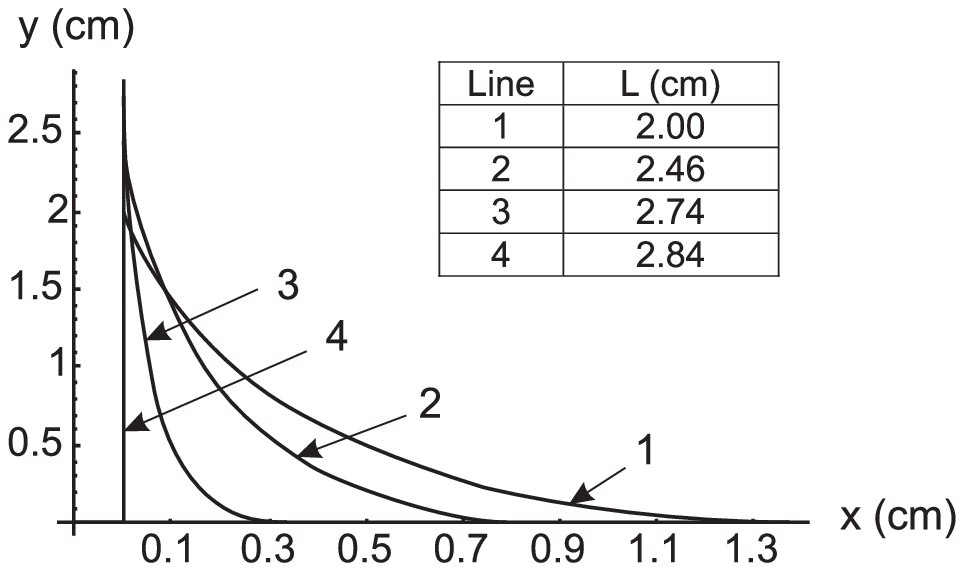}}}
  \subfigure[Varying $\mu$.]
  {\label{fig7:2}\scalebox{0.6}{\includegraphics{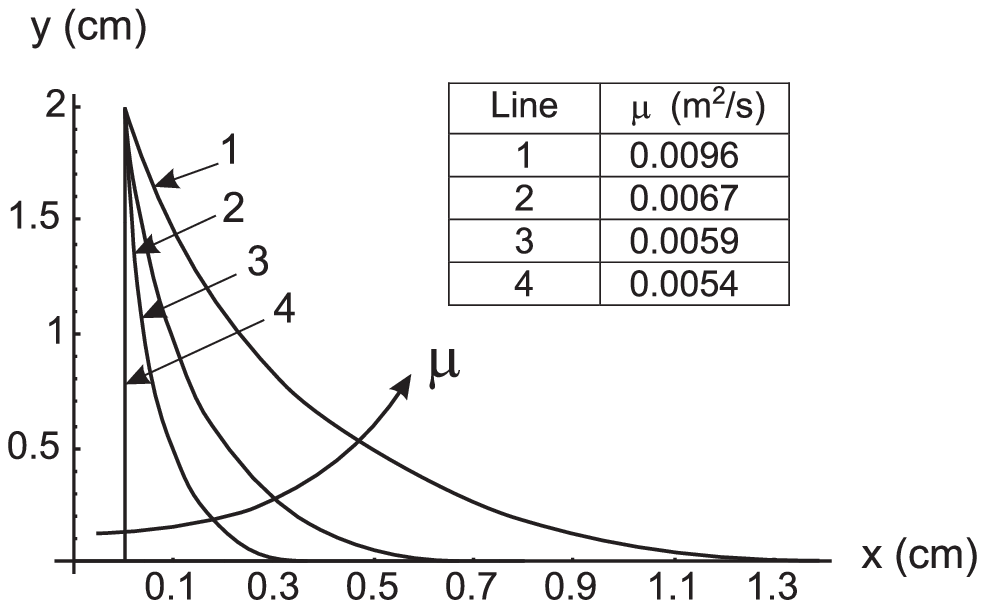}}}\\
  \subfigure[Varying $\vb$.]
  {\label{fig7:3}\scalebox{0.6}{\includegraphics{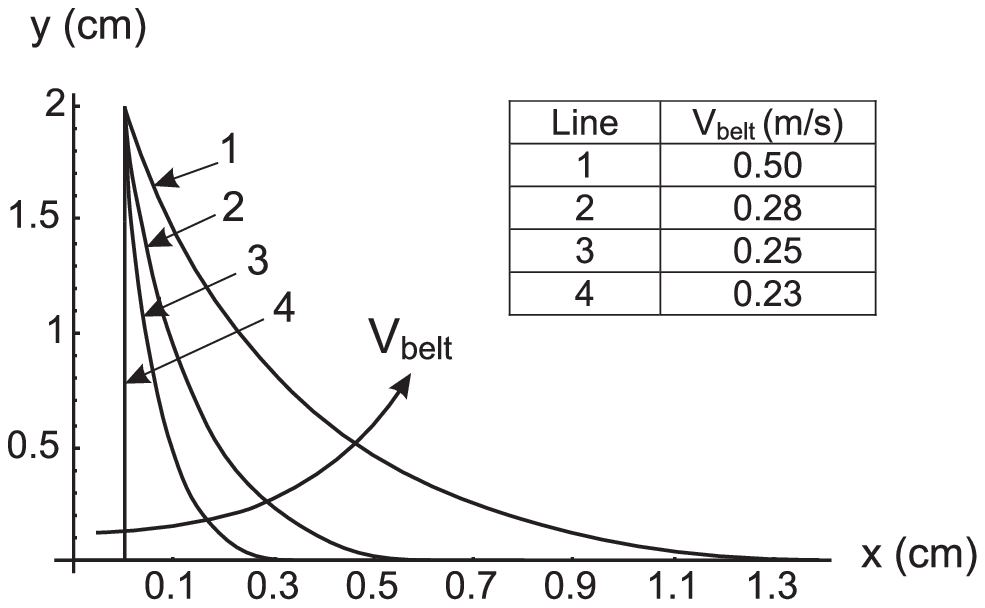}}}
  \subfigure[Varying $\vn$.]
  {\label{fig7:4}\scalebox{0.6}{\includegraphics{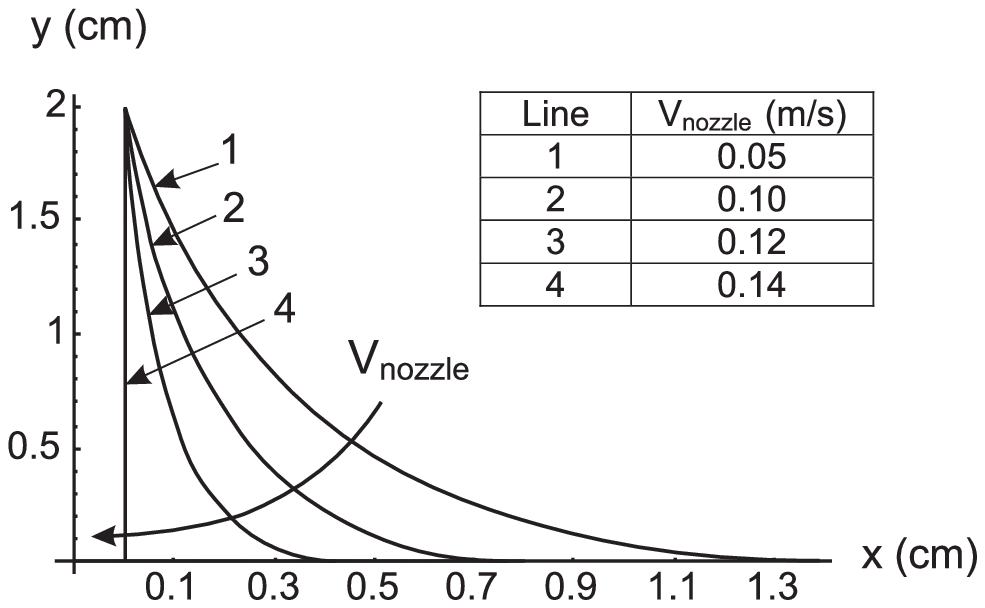}}}
  \caption{The shapes of a jet for different values of the process parameters $(L,\mu,\vb,\vn)$. The reference values for the parameters are $\eta=3.2\,\mathrm{Pa\, s}$, $\rho=1000\,\mathrm{kg/m^3}$, $L=2\cdot10^{-2}\,\mathrm{m}$, $\vb=0.5\,\mathrm{m/s}$, and $\vn=0.05\,\mathrm{m/s}$.}
  \label{fig7}
\end{figure}
In Fig.~\ref{fig7}, we present the shapes of the jet for specific values of the parameters along each of these curves.
\par
Summarizing the numerical experiments, we observe that by increasing the flow velocity at the nozzle or the distance between the belt and the nozzle the jet shape becomes more vertical; the same is true if we decrease the velocity of the belt or the kinematic viscosity. The jet becomes exactly vertical when the parameter point $(A,\vn,\B)$ approaches the critical surface $\{\I(0;A,\vn)=\B\}$.
\section{Conclusions}  \label{sec:concl}
In this paper we propose a mathematical model of the falling of a viscous jet onto a moving surface. We assume that the jet is falling under gravity and has a curved shape. The model consists of two differential equations, one for the flow velocity and one for the angle describing the jet's shape. An additional relation fixes the unknown length of the jet.
\par
The initial system of equations is partially solved and then transformed to a first-order differential equation for the velocity. By introducing an additional scalar parameter $w$ the problem is reformulated as an algebraic equation for $w$ (\ref{eq:for_w}). For this equation we formulate an existence condition (\ref{eq:exisCond}) and prove uniqueness, thus giving a complete characterization of existence and uniqueness for the original equations.  Finally, we solve the equation for $w$ numerically and recover the solution of the original problem.
\par
We have shown that if the existence condition (\ref{eq:exisCond}) is satisfied, then the shape of the jet is curved; we conjecture that in the alternative case the jet is vertical, but this case lies outside of the scope of this paper. Furthermore, the model shows that the curved jet becomes more vertical when: i) the distance between the nozzle and the surface increases, ii) the flow velocity at the nozzle increases, iii) the surface velocity decreases, or iv) the kinematic viscosity of the fluid decreases. These results correspond with those observed in the basic experiment described in the introduction.
\Appendix
\section{Calculation of $\I(0;A,\vn)$} \label{sec:Appendix}
First we calculate $v(\tm;0)$ analytically. The differential equation for $v(\tm;0)$ follows from (\ref{EqForV(t)1}) and (\ref{EqForV(t)2}),
\be
 v'(\tm;0)=-v^2(\tm;0)(A\tm+v(\tm;0)), \ \ \
 v(0;0)=1.\label{EqForV_Wis0}
\ee
By replacing $v(\tm;0)$ by $Z(\tm)=1/v(\tm;0)$, we find
\be
 Z'(\tm)Z(\tm)=AZ(\tm)\tm+1, \ \ \ Z(0)=1.\label{EqForZ}
\ee
We seek for a solution of (\ref{EqForZ}) in parametric form. With the substitution
\be
  Z(z) =z  + A/2\tm^2(z),\label{ParRepl}
\ee
where $z$ is a parameter, (\ref{EqForZ}) becomes
\be
 \tm'(z)=A\tm^2(z)+z, \ \ \ \tm(1)=0.\label{ParEqForTau}
\ee
Here the initial condition is deduced from (\ref{ParRepl}) by setting $\tm(z)=0$ and  $Z(z)=1$. This differential equation is known  as the special Riccati equation~\cite[p.~4, type~4]{Zaitsev} and has the solution
\be
 \tm(z)=\frac{\sqrt{2z}\left(\Jtt\left(\frac{\sqrt{2A}z^{3/2}}{3}\right)c_1
-\Jmtt\left(\frac{\sqrt{2A}z^{3/2}}{3}\right) \right)}
{\sqrt{A}\left(\Jot\left(\frac{\sqrt{2A}z^{3/2}}{3}\right)
+\Jmot\left(\frac{\sqrt{2A}z^{3/2}}{3}\right)c_1\right)},
\label{SolforTau1}
\ee
with
\be
  c_1=\frac{\Jmtt(\alpha)}{\Jtt(\alpha)},\
  \ \alpha=\sqrt{2A}/3. \label{SolforTauConst1}
\ee
Here the functions $J_\alpha$ are the Bessel functions of the first kind. The velocity $v$ is given by
\be
 v(z)=\frac{1}{z}\left(1+\left(
 \frac{\Jtt\left(\frac{\sqrt{2A}z^{3/2}}{3}\right)c_1
-\Jmtt\left(\frac{\sqrt{2A}z^{3/2}}{3}\right) }
{\Jot\left(\frac{\sqrt{2A}z^{3/2}}{3}\right)
+\Jmot\left(\frac{\sqrt{2A}z^{3/2}}{3}\right)c_1}\right)^2\right)^{-1}.
\label{SolforV1}
\ee
To write the result in a more elegant form we replace the parameter $z$ by $\tilde{z}=\sqrt{2A}z^{3/2}/3$ (we then omit tildes)
\begin{eqnarray}
 \tm(z)=\frac{(6z)^{1/3}}{A^{2/3}}
 \frac{\Jtt(z)c_1
-\Jmtt(z)} {\Jot(z) +\Jmot(z)c_1},
\label{SolforTau2} \\
 v(z)=\frac{(2A)^{1/3}}{(3z)^{2/3}}\left(1+\left(\frac{\Jtt(z)c_1
-\Jmtt(z)} {\Jot(z) +\Jmot(z)c_1}\right)^2\right)^{-1}.
\label{SolforV2}
\end{eqnarray}
To calculate $\tme(0)$ from the solution (\ref{SolforTau2}) and (\ref{SolforV2}) it is necessary to find $z^*$ satisfying
\be
 \vn=\frac{(2A)^{1/3}}{(3z^*)^{2/3}}\left(1+\left(\frac{\Jtt(z^*)c_1
-\Jmtt(z^*)} {\Jot(z^*) +\Jmot(z^*)c_1}\right)^2\right)^{-1},
\label{SolforZstar}
\ee
and then substitute $z=z^*$ into (\ref{SolforTau2}).
\par
The equation (\ref{SolforZstar}) has many solutions. A correct solution $z^*$ is the first solution of (\ref{SolforZstar}) after the point $\alpha$. It is convenient to search for $z^*$ in the interval $(\alpha,z_0)$ using the bisection method \cite{Numerical}. Here, $z_0$ is the first zero of $v(z)$ according to (\ref{SolforV2}) after the point $\alpha$.
\par
Next, we have to find a correct $z_0$. Because zeros of $v(z)$ coincide with zeros of
\be
 \Jot(z) +\Jmot(z)c_1,  \label{eq:fz0Step1}
\ee
we can look for the first zero of (\ref{eq:fz0Step1}) after  $\alpha$. Using (\ref{SolforTauConst1}), we can rewrite the latter as
\be
 \Jot(z_0)\Jtt(\alpha) +\Jmot(z_0)\Jmtt(\alpha)=0.  \label{eq:fz0Step2}
\ee
This equation can be rewritten in terms of Airy functions \cite[10.4.22 and 10.4.27]{Abram} as
\be
\ \ Bi(-\hzz)Ai'(-\halp)-Ai(-\hzz)Bi'(-\halp)=0, \
\hzz=\left(\frac{3z_0}{2}\right)^{\frac{2}{3}}, \
\halp=\left(\frac{3\alpha}{2}\right)^{\frac{2}{3}}.
\label{eq:fz0Step3}
\ee
Using the representation of Airy functions via modulus and phase \cite[10.4.69 and 10.4.70]{Abram}
$$
Ai(-\hzz)=M(\hzz) \cos \theta(\hzz), \ \
Bi(-\hzz)=M(\hzz)\sin \theta(\hzz),
$$
$$
Ai'(-\halp)=N(\halp) \cos \phi(\halp), \ \
Bi'(-\halp)=N(\halp)\sin \phi(\halp),
$$
we see that (\ref{eq:fz0Step3}) becomes
\be
 \sin(\theta(\hzz)-\phi(\halp))=0. \label{eq:fz0Step4}
\ee
For large $\hzz \gg 1$ and  $\halp \gg 1$ the asymptotic expressions for $\theta(\hzz)$ and $\phi(\halp)$ \cite[10.4.79 and 10.4.81]{Abram} are given by
$$
\theta(\hzz)=\frac{\pi }{4}-\frac{3}{2} \hzz^{2/3}
\left(1-\frac{5}{32 \hzz^3}+\frac{1105}{6144 \hzz^6}+...\right),
$$
and
$$
\phi(\halp)=\frac{3 \pi }{4}-\frac{3}{2}\halp^{2/3}
\left(1+\frac{7}{32 \halp^3}-\frac{1463}{6144 \halp^6}+...\right),
$$
or in terms of $z_0$ and $\alpha$ (\ref{eq:fz0Step3})
\be
\theta(z_0)=\frac{\pi }{4}-z_0\left(1-\frac{5}{72
z_0^2}+\frac{1105}{31104 z_0^4}+...\right), \label{eq:fz0Step5}
\ee
and
\be
\phi(\alpha)=\frac{3 \pi }{4}-\alpha\left(1+\frac{7}{72
\alpha^2}-\frac{1463}{31104 \alpha^4}\right). \label{eq:fz0Step6}
\ee
After substituting (\ref{eq:fz0Step5}) and (\ref{eq:fz0Step5}) into (\ref{eq:fz0Step4}) for $\alpha\gg 1$ we find
\be
 z_0\approx \alpha+\pi/2.
\ee
When $\alpha$ is not large we can find $z_0$ numerically by looking for a solution of (\ref{eq:fz0Step2}) in the interval $(\alpha,\alpha+\pi)$.
\par
Once $z_0$ is found we find $z^*$ and consequently compute $\tme(0)$. Knowing $\tme(0)$ we can compute $I(0)$. To avoid computation of the integral
$$
\I(0)=\int_0^{\tme(0)} v(\tm;0)\,d\tm,
$$
we can calculate this integral using the differential equation (\ref{EqForV_Wis0}), when written as
\be
  \left(\frac{1}{v(\tm;0)}\right)'=A\tm+v(\tm;0). \label{eq:apvj1}
\ee
By integrating this equation  from $0$ to $\tme(0)$, we get
\be
  \left(\frac{1}{v(\tme(0);0)}-\frac{1}{v(0;0)}\right)=A\frac{\tme(0)^2}{2}+
  \int_0^{\tme(0)} v(\tm;0)\,d\tm. \label{eq:apvj2}
\ee
We use the definitions of $I(0)$ and $\tme(0)$, together with the initial condition $v(0;0)=1$ to obtain
\be
 \I(0)=\frac{1-\vn}{\vn}-A\frac{\tme(0)^2}{2}.
\ee

\bibliographystyle{plain}
\bibliography{ArtBib}

\end{document}